\documentstyle[prb,twocolumn,aps]{revtex}
\input epsf
\begin{document}
\draft
\title{Coupling strength of charge carriers to spin fluctuations in
high-temperature superconductors}
\author{J.P. Carbotte$^\star$, E. Schachinger$^\dagger$ \& D.N. Basov$^\ddag$}
\address{$^\star$Department of Physics and Astronomy, McMaster University\\
  Hamilton, Ont. L8S 4M1, Canada
}
\address{$^\dagger$Institut f\"ur Theoretische Physik, Technische Universit\"at
Graz\\A-8010 Graz, Austria}
\address{$^\ddag$Physics Department,
     University of California - San Diego\\
     La Jolla California 92093-0319,
     U.S.A.}
\date{\today}
\maketitle
\newpage
{\bf In conventional superconductors, the most direct evidence of the mechanism responsible for
superconductivity comes from tunnelling experiments in which a clear image of the
electron-phonon interaction is revealed.\cite{McMillan,Carb1} The observed structure in the
current voltage characteristics at the phonon energies can be used to measure, through
inversion of the Eliashberg equations, the electron phonon spectral density
$\alpha^2F(\omega)$.\cite{McMillan} The coherence length in conventional materials is
long and the tunnelling process probes several atomic layers into the bulk of the
superconductor. On the contrary, in the high $T_c$ oxides, particularly
for $c$-axis tunneling, the coherence length can be quite short
and in an optical experiment or in neutron
scattering experiments the bulk of the sample is probed. Therefore, these
spectroscopies become the methods of choice for the investigation of mechanisms
of high-$T_c$ superconductivity. Accurate reflectance measurements in the
infrared range and precise polarized neutron scattering data are available
for a variety of oxides.\cite{Mars,Puchk,Bourg2} In this paper we show that
conducting carriers studied by means of infrared spectroscopy reveal strong
coupling to a resonance structure in the spectrum of spin fluctuations
examined with neutron scattering. The coupling strength inferred from
experiment is sufficient to account for high values of $T_c$ which signals
the prominent role of spin excitations in the superconductivity of oxides.}

\begin{figure}[t]
\vspace*{-0.5cm}
\hspace*{1cm}
\epsfxsize=6cm
\epsffile{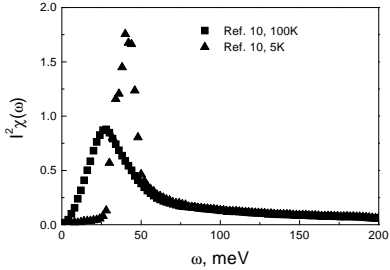}
\caption{Spin polarized inelastic neutron scattering data for YBa$_2$Cu$_3$O$_{6.92}$.
Neutron 
scattering results taken at 100 K in the normal state (solid squares) compared with similar 
data at 5 K in the superconducting state (solid triangles) showing the $41\,$meV
resonance. The data have been scaled by a constant coupling strength $I^2$
fixed to get a $T_c = 100\,$K.}
\label{f1}
\end{figure}

There have been many suggestions as to the mechanism involved in the superconductivity
of the oxides.  While, so far, no consensus has yet emerged, the state itself is widely accepted
to have $d$-wave symmetry \cite{Scal,Mont,Woll,Tsue} with the gap on the Fermi surface vanishing
along the main diagonals of the two-dimensional Brillouin zone. In YBCO there also exists 
extensive spin polarized  inelastic neutron scattering data. These experiments reveal
that spin excitations persist on a large energy scale,\cite{Bourg1} over several 100 meV,
but are mainly confined around the $(\pi,\pi)$-point in momentum space.  Also, in the
superconducting state, a new peak emerges out of, or is additional to, the spin excitation
background which is often referred to as the $41\,$meV resonance\cite{Bourg2} (Fig.~\ref{f1}).
 This peak has
received much attention but its origin remains uncertain.\cite{Bulut,Demler}
In one view, it is due to a readjustment in the spin excitation spectrum due to 
the onset of superconductivity.\cite{Bulut} Such a readjustment of spectral weight
with a reduction below twice the superconducting gap value $\Delta_0$
\cite{Nuss,Varma} is expected on general ground and is generic to electronic mechanisms.
A second view is that it is a resonance in the $SO(5)$ unification\cite{Demler} of magnetism
and superconductivity.

If the spin excitations are strongly coupled to the charge carriers they should
be seen in optical experiments.
The normal state optical conductivity $\sigma(\omega)$ as a function of frequency
$(\omega)$ depends on the electron self energy $\Sigma(\omega)$ which describes the effect
of interactions on electron motion. In an electron-phonon system the
electron-phonon
interaction spectral density, $\alpha^2F(\omega)$, is approximately\cite{Mars}
(but not exactly) equal to $W(\omega)$, a second derivative of the inverse
of the normal state optical conductivity
\begin{equation}
  \alpha^2F(\omega) \approx W(\omega) = {1\over 2\pi}{d^2\over d\omega^2}\left[
    \omega\Re{\rm e}{1\over\sigma(\omega)}\right].
  \label{e1}
\end{equation}
In the phonon energy range, the correspondence is remarkably close and determines
$\alpha^2F(\omega)$ with good accuracy.  At higher energies, additional, largely negative
wiggles come into $W(\omega)$ which can simply be ignored as they are not part of 
$\alpha^2F(\omega)$.  Note that (\ref{e1}) is dimensionless and so determines the absolute
scale of the electron-phonon interaction spectral density as well as its shape in frequency.
This fact is important as it allowed Marsiglio {\it et al.}\cite{Mars} to determine
the $\alpha^2F(\omega)$ of K$_3$C$_{60}$ from its
optical conductivity by inversion (\ref{e1}) and to conclude from a solution
of the Eliashberg equations that it is large enough to explain the observed
value of critical temperature. ($T_c$ is related to the mass enhancement
factor $\lambda$, twice the first inverse moment of $\alpha^2F(\omega)$.\cite{Carb1})

The formalism for the normal state conductivity can also be applied to spin excitations.
If we ignore anisotropy as a first approximation, we can proceed by introducing an
electron-spin excitation spectral density denoted by $I^2\chi(\omega)$ with its scale set 
by the coupling strength to the charge carriers, $I^2$, and $\chi(\omega)$ the imaginary part
of the spin susceptibility measured in spin
 \noindent
polarized  
inelastic neutron scattering experiments
averaged over all momentum in the Brillouin zone. At low temperatures $\chi(\omega)$
contains the $41\,$meV resonance observed in the superconducting state. To illustrate our
main point we will use here $\chi(\omega)$
directly from experimental results on a YBa$_2$Cu$_3$O$_{6.92}$ sample with $T_c =91\,$K, near
optimum doping and for which results exist at the temperatures $T=100\,$K and $T=5\,$K,
\cite{Bourg1}  both properly calibrated in units of $\mu_B^2/eV$
($\mu_B$ is the Born magneton) as shown in Fig.~\ref{f1}. We multiply $\chi(\omega)$ at
 $T=100\,$K by a constant
 coupling  $I^2$ fixed to get $T_c = 100\,$K.\cite{Schach}
The mass enhancement factor $\lambda$ (twice the first inverse moment of
$I^2\chi(\omega)$) obtained is 2.6 and is, to within 10 percent, the same as obtained
from the $W(\omega)$ derived from the normal state experimental data of
Basov {\it et al.}\cite{Basov} in YBa$_2$Cu$_3$O$_{6.95}$ and from our calculated 
$W(\omega)$ at $T_c$. In a preliminary attempt to invert Collins {\it et
  al.}\cite{Coll} found a $\lambda$ of three which is greater than our value.
Their twinned crystals
exhibited a higher optical scattering rate than our untwinned crystal and
consequently they obtained about 50\% more weight in the main peak of
$I^2\chi(\omega)$ around $30\,$meV.

\begin{figure}[t]
\vspace*{-0.7cm}
\hspace*{1cm}
\epsfxsize=6cm
\epsffile{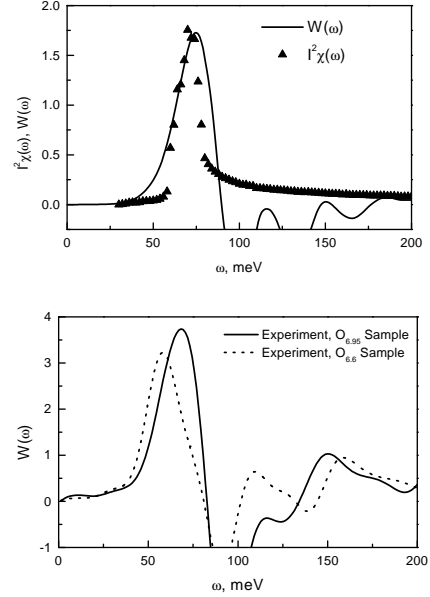}
\caption{Inversion of the superconducting state optical conductivity.
Results (top frame) for $W(\omega)$ (solid curve) with $\sigma(\omega)$ calculated in the
 superconducting state at 5 K and based on the
spectral density $I^2\chi(\omega)$ (solid triangles) equal to the $5\,$K measured neutron data.
In the region of the main resonance peak, the agreement with the 
input data, which has been shifted by the gap value (solid triangles), is excellent
as to height and width.  At higher energies, wiggles appear in $W(\omega)$ which are
not present in $I^2\chi(\omega)$. Bottom frame,
$W(\omega)$ derived from the experimental data for the conductivity of an optimally
doped YBa$_2$Cu$_3$O$_{6.95}$ single crystal with $a$-polarization (solid curve) and for
an underdoped, untwinned YBa$_2$Cu$_3$O$_{6.6}$ single crystal (dashed line).}
\label{f2}
\end{figure}

In order to access lower temperatures, we need to understand how the $I^2\chi(\omega)$ structure
enters the superconducting state optical conductivity. To this end, we have
done a series of
calculations of the superconducting state $\sigma(\omega)$ for a $d$-wave superconductor
including inelastic scattering. Details have been presented in the work of
Schachinger {\it et al.}\cite{Schach} We used their prescription to calculate the
theoretical $\sigma(\omega)$ using the neutron data taken for
YB$_2$C$_3$O$_{6.92}$ (at $5\,$K) as $\chi(\omega)$
multiplied by the same value of the coupling strength $I^2$ previously
determined to obtain a
$T_c$ of $100\,$K from the normal state neutron data. We then inverted
this theoretical $\sigma(\omega)$ data using
Eq.~(\ref{e1}). The result of this inversion is compared in the top frame of Fig.~\ref{f2}
(solid line) with our input spectral density $I^2\chi(\omega)$ (solid triangles) shifted
in energy by the gap $\Delta_0 = 27\,$meV of our theoretical calculations.\cite{Mars2}

The absolute scale of $I^2\chi(\omega)$ in the resonance region
is well given by the peak value in the solid curve. This peak is followed by negative
wiggles which are not in the
 \noindent
 original input spectrum because $W(\omega+\Delta_0)$ is not exactly
$I^2\chi(\omega)$. Nevertheless, such a procedure allows us to see quite directly
by spectroscopic means some of the features of $I^2\chi(\omega)$ and, more importantly gives us
information on its absolute value at maximum.  The long tails in $I^2\chi(\omega)$ at higher
energy extending well beyond the resonance are not resolved in $W(\omega)$ but cause
$\tau^{-1}(\omega)$, defined as
$\Re{\rm e}\{\sigma^{-1}(\omega)\}$, to rise in a quasi linear fashion 
 at high
frequencies\cite{Puchk} in both normal and superconducting state, as is observed. This 
quasilinear rise was the motivation for the marginal Fermi liquid
model \cite{Varma} which gives $\tau^{-1}(\omega)\propto\omega$ and a constant spectral
density for $\omega > T$ extending to high energies. If we approximate the normal state
experimental $\tau^{-1}(\omega)$ data\cite{Puchk} at $T_c$ by a straight line for
$0\le\omega\le 200\,$meV, 
we get 0.3 as the
weight of the spectral density for all frequencies  $\omega > T$ and a $\lambda$ of
2.8 quite consistent with our previous estimates. It is important to
realize that, in as much as the $41\,$meV resonance is near
 $2\Delta_0$, the density of quasiparticle states (not shown here), 
has structure at $3\Delta_0$ in our calculations, a
well established
feature of tunnelling data particularly in Bi2212.\cite{DeWilde}

In the bottom frame of Fig.~\ref{f2} we show experimental results obtained from the data by
Basov {\it et al.}\cite{Basov} on application of the prescription (\ref{e1}) to $a$-axis
conductivity data on an
untwinned single crystal.  The $41\,$meV resonance is clearly resolved as a peak at approximately
69 meV in the solid curve (the gap is $27\,$meV).  The height of this peak is about 3 and gives an
absolute measure of the coupling between charge carriers and spin
excitations. On comparison with the top frame of Fig.~\ref{f2}, we see that the coupling to
the $41\,$meV resonance is larger in the experiment 
than the value assumed in the calculations
that generated the theoretical results of that frame. This is not surprising since
we have used the spin polarized inelastic neutron scattering data set measured on a
near optimum $91\,$K sample of YBa$_2$Cu$_3$O$_{6.92}$ while the neutron results for slightly
overdoped YBCO are very different\cite{Bourg2} although the $T_c$ value is hardly
affected. This large dependence of $\chi(\omega)$ on the sample
can be used to argue against their role in establishing $T_c$. However,
the function that controls the conductivity is a complicated weighting of the
spin susceptibility involving details of the Fermi surface and points in the
Brillouin zone away from $(\pi,\pi)$ as well as the coupling to the charge
carriers. Thus, the correspondence between $I^2\chi(\omega)$ and $\chi(\omega)$
is complicated. What optical experiments reveal is that $I^2\chi(\omega)$ is not as strongly
dependent on doping as is $\chi(\omega)$.

In Fig.~\ref{f2}, bottom frame, we present experimental
results for $W(\omega)$ in underdoped, untwinned YB$_2$C$_3$O$_{6.6}$ (dashed line)
 and compare with the
optimally doped case (solid line). It is interesting to note that the peak in the underdoped
case is slightly reduced in height reflecting a reduction in $T_c$.  It is also shifted to 
lower energies.  Some experiments \cite{Ponom} indicate a reduction in gap value with 
underdoping in YBCO while many experiments show an important increase in Bi2212.\cite{DeWilde}
Even if the gap is assumed to stay the same at $27\,$meV, the spin polarized neutron resonant
frequency is known to decrease with doping.\cite{Fong1} Accounting for this gives almost 
exactly the downward shift observed in our experimental data of Fig.~\ref{f2} (bottom
frame).

Very recently, inelastic neutron scattering data in Bi2212\cite{Fong2} have been
published.  They show a resonance peak at $43\,$meV in the superconducting state and establish a
similarity with the earlier results in YBCO. We have inverted the optical data of Puchkov 
{\it et al.} \cite{Puchk} in this case and find that 
coupling at low temperatures to the observed superconducting state
spin resonance peak is a general phenomena in both YBCO and Bi2212.

Spin excitations are seen in an appropriately chosen second derivative of the
superconducting state optical conductivity and the strength of their coupling
to the charge carriers determined from such data. The coupling to the excitations
including the $41\,$meV resonance is large enough in YBCO that it can account for
superconductivity at that temperature. At $T_c$ the spectrum obtained from
experiment\cite{Puchk} gives a value of the mass enhancement parameter
$\lambda$ which is close to the value used in our model calculations to
obtain a critical temperature of $100\,$K.

{\bf Acknowledgments.}
     This work was supported in part by the Natural Sciences and Engineering Research Council
of Canada (NSERC), the Canadian Institute for Advanced Research (CIAR).
Work at UCSD is supported by the NFS ``Early Career Development'' program.
DNB is a Cottrel Scholar of the Research Corporation.
We thank J.E.~Hirsch, P.B.~Hirschfeld, P.B.~Littlewood, F.~Marsiglio, E.J.~Nicol, D.~%
Scalapino, T.~Timusk, and I.~Vekhter for interest.
\end{document}